Vishal Kaushik[1], Swati Rajput[1], Sulabh Srivastav[1], Lalit Singh[1], Prem Babu[1], Elham Heidari[4], Moustafa Ahmed[5], Yas Al-Hadeethi[5], Hamed Dalir[2,4], Volker J. Sorger[2] and Mukesh Kumar[1,3]

1Optoelectronic Nanodevice Research Laboratory, Department of Electrical Engineering, Indian Institute of Technology (IIT) Indore, India
2Department of Electrical and Computer Engineering, George Washington University, USA
3Centre for Advanced Electronics (CAE), Indian Institute of Technology (IIT) Indore, India
4Optelligence LLC, Alexandria, VA 22302, USA
5Physics Department, Faculty of Science, King Abdulaziz University, P.O. Box-80207, Jeddah, 21589, Saudi Arabia
Corresponding email addresses: hdalir@gwu.edu, mukesh.kr@iiti.ac.in

# On-chip Nanophotonic Broadband Wavelength Detector with 2D-Electron Gas

Subtitle: Nanophotonic Platform for Wavelength Detection in Visible Spectral Region

**Abstract:** Miniaturized, low-cost wavelength detectors are gaining enormous interest as we step into the new age of photonics. Incompatibility with integrated circuits or complex fabrication requirement in most of the conventionally used filters necessitates the development of a simple, on-chip platform for easy-to-use wavelength detection system. Also, intensity fluctuations hinder precise, noise free detection of spectral information. Here we propose a novel approach of utilizing wavelength sensitive photocurrent across semiconductor heterojunctions to experimentally validate broadband wavelength detection on an on-chip platform with simple fabrication process. The proposed device utilizes linear frequency response of internal photoemission via 2-D electron gas in a ZnO based heterojunction along with a reference junction for coherent common mode rejection. We report sensitivity of 0.96 µA/nm for a broad wavelength-range of 280 nm from 660-940 nm. Simple fabrication process, efficient intensity noise cancelation along with heat resistance and radiation hardness of ZnO makes the proposed platform simple, low-cost and efficient alternative for several applications such as optical spectrometers, sensing and IOTs.

## 1 Introduction:

As electronics-based technologies are reaching their performance bottlenecks, research community is shifting their focus from electrons to photons. Photons enjoy various advantages over the electrons such as wider bandwidth higher speeds [1-4]. Integrated photonics has witnessed immense interest as it carries the promise of significant reduction of size, weight, manufacturing costs, and power consumption while improving reliability, in comparison to the assembling and packaging of multiple discrete photonics components [5-10]. Wavelength sensitive detectors with low intensity noise is a technology that is used in several fields of research or engineering where spectral information is essential. Hence, low cost, on-chip wavelength detectors which can be extremely useful in the fields of sensing, IOTs, communication and environmental monitoring have attracted the attention of the researchers in recent years. Thus, enormous amount of research efforts has been concentrated on the development of different types of spectral filters such as dispersive and absorptive spectrum filters. On one hand, dispersive filters utilize optical interference owing to difference between optical path-length for different wavelengths, some examples of such filters include Fabry-Perot cavities, gratings and photonic crystals. In order to achieve the desired optical path-difference bulky dispersive filters are typically required. Such systems are not compatible with on-chip photonics, or when the system is Size, Weight and Power (SWaP) constrained. In addition to this, strong dependence on the angle of incident light is another limitation for such system as it requires the use of collimation optics, further adding to bulk of the system. On the other hand, semiconductor-based filters offer on-chip compatible alternative which relies on materials with wavelength selective absorption or resonant structure such as photonic crystal cavities. Nevertheless, to achieve high spectral resolution it is inevitable to integrate complex geometries with very low fabrication tolerance, additionally broadband operation is achieved by employing multiple layers of different materials. This not only increases the complexity but also leads to excessive production costs [14,15]. Another approach that uses 2D materials is gaining a lot of attention due to their unique opto-electronic properties such as ultrahigh responsivity, fast photoresponse and broad detection range. However, issues such as lack of maturity in large scale growth of high-quality film along with concerns related with pattering and highly quality ohmic contacts damages the wide appeal that 2DM enjoy. In addition to this most of the devices are usually susceptible to atmospheric noise due to linearity constrains in amplitude domain. Internal photoemission in Metal-Semiconductor Schottky junctions has drawn a lot of interest owing to its capability for photodetection in sub-bandgap regime [16-19]. Frequency dependence of Quantum Yield (Y) in such devices has been thoroughly studied and shown to be dependent on frequency with a relation $Y \propto (h\nu - \phi_\beta)^\gamma$ where $\phi_\beta$ is the potential barrier at the interface and $\gamma$ is number ranging from 1 to 3 [19,20]. Internal photoemission display near linear frequency response with respect to frequency for a wide range of spectrum provided $\gamma$ for the system is close to 1. These structures suffer from low responsivity due to poor internal quantum efficiency [21-23]. Lately introduction of 2D materials with plasmonic mode is being actively pursued to enhance the responsivity to several orders of magnitude however, the associated metallic losses and momentum mismatch severely hinders their application [24-28].

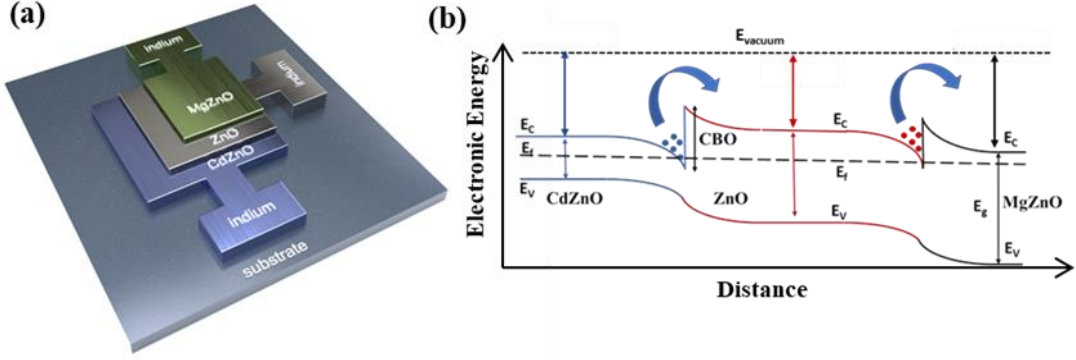

Fig. 1 (**a**) Schematic of the proposed nanophotonic device for wavelength detection. Active layers CdZnO-ZnO-MgZnO forming two heterojunctions at the interface of CdZnO-ZnO and ZnO-MgZnO along with individual contact pads enabling application of bias voltage. (**b**) Electronic band diagram of the double heterojunction along with confined charge carriers. Horizontal axis represents distance from the substrate along the thickness of the deposited layers and vertical axis represents the electronic energy. Conduction band offset at both the junctions are 0.54 and 0.56 eV.

Recently sub-bandgap photodetection utilizing Two-Dimensional Electron Gas (2-DEG) in semiconductor heterojunction has been shown to exhibit high responsivity where high density of 2-D confined electrons absorbs photons energy to jump across the Conduction Band Offset (CBO) at the heterojunction as shown in Fig. 1(b). Wide bandgap semiconductors like ZnO offer higher barrier height useful for applications in visible and IR region [29-33]. Zinc Oxide exhibits bandgap engineering when alloyed with MgO and CdO. Its bandgap can be increased or decreased with increasing percentage of MgO and CdO respectively [34-37]. Bandgap engineering in ZnO based alloy has been used to demonstrate heterojunctions with very high density of 2-DEG confined at the junction barrier. The electronic band diagram of the heterojunctions is shown in Fig. 1(b) where high density of confined electrons is present on both junction barriers. With applied bias across the junction the current is restricted due to potential barrier present in the conduction band. However, as light is incident at the junction electrons gains sufficient energy to overcome the barrier and drive photocurrent as shown in fig. 1(b). Tightly confined electron gas exhibited in the proposed ZnO based heterojunctions shows low losses while offering superior momentum matching which paves the way for low reflection as compared to their metal counter parts. ZnO-heterojunction based platform with high density of 2-DEG on one hand, offers superior Internal Quantum Efficiency (IQE) over traditional emitters like metals [34,37]. On the other hand, it also displays almost linear relationship between photocurrent and photonic energy for a broad range of frequency spectrum. Since the current is dependent on two variables i.e., frequency and amplitude, two such heterojunctions must be employed to distinguish and eliminate the contributions of light intensity for error free wavelength detection.

We report (in proof-of-principle) a broadband, on-chip optical wavelength detector in part visible part IR region with potential applications in the fields of optical spectrometer, sensing and demodulation of wavelength modulated optical signals. The proposed approach utilizes multiple asymmetric heterojunctions of ZnO based material system to achieve wavelength detection of the incident light. Thermal resistance and radiation hardness of ZnO makes the device suitable for operation in harsh environment. Additionally, the device is engineered to reduce intensity dependence of the photoemission current via 2-DEG formed at the ZnO based heterojunctions. Linear dependence on frequency, simple fabrication process along with low dependence on intensity of the light makes the proposed approach very simple and effective for reducing intensity-based noise. Where most platforms use different materials or resonators for targeting different part of the spectrum, the proposed approach uses bandgap engineering with same material system. Present work employs three different bandgap materials (i.e., CdZnO-ZnO-MgZnO) forming two heterojunctions.

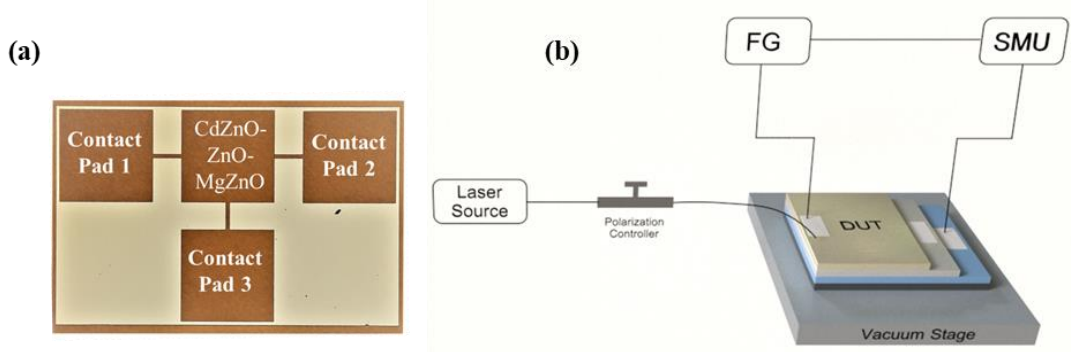

Fig. 2 (a) Optical image of the fabricated device. CdZnO-ZnO-MgZnO layer deposited over each other and connected to an individual contact pad. (b) Setup for opto-electronic characterization of the Device Under Test (DUT). Both the junctions are biased using Signal Measurement Unit (SMU) and Function Generator (FG) and optically excited with photonic energies from 0.8 to 1.87 eV using different laser source.

Materials have been carefully engineered to reduce intensity dependence from the current difference of both the junctions. This difference current display reduced intensity dependence along with maintaining linear frequency response. A particular multiplication factor is introduced to further reduce the intensity dependence of the device. This could pave the way for relatively simple, on-chip, low noise wavelength detection. Transparent nature of 2-DEG on one hand ensures equal exposure of light over both the junctions and thus provides a spatially and temporally coherent reference photocurrent which is used to compensate for the intensity noise. On the other hand, high density of 2-DEG offers high IQE along with high photocurrent making it a perfect platform for applications in wavelength demodulation, sensing, IOT etc.

## 2  Device design:

The schematic diagram of the device is shown in Fig. 1(a), where three active layers of ZnO based compound semiconductor alloy (CdZnO-ZnO-MgZnO) are shown. Initially a layer of $Cd_{0.3}Zn_{0.7}O$ with thickness and band gap of 100 nm and 2.78 eV respectively, is deposited over silicon wafer using Dual Ion Beam Sputtering (DIBS) system. The thickness and band gap of the layer is experimentally confirmed with the help of Atomic Force Microscopy (AFM) and UV visible spectroscopy respectively. The deposited layer is then patterned with the help of direct UV laser writer PICOMASTER 100 and the structure is subsequently wet etched in diluted HCl solution. Similarly, 100 nm thick layers of ZnO and $Mg_{0.25}Zn_{0.75}O$ with bandgaps of 3.36 eV and 4 eV respectively are deposited, patterned and etched to obtain the final device. The deposited layers form two heterojunctions of CdZnO-ZnO and ZnO-MgZnO. Patterning is done in a way so that each layer is connected to individual electrodes which allow collection of photocurrents through each junction. The conduction band offset for these heterojunctions can be estimated with the formula CBO = 0.9*ΔEg, where ΔEg is the bandgap difference of the materials [33,34]. CBO of 0.54 eV and 0.56 eV is estimated using the above criteria while, the presence of confined charge carriers with a density of $2.3 \times 10^{20}$ and $1.8 \times 10^{20}$ respectively is experimentally confirmed with help of capacitance-voltage graph. Fig. 2(a) shows the optical image of the fabricated device. The active region of the device consists of the layers of CdZnO-ZnO-MgZnO which are deposited one over the other forming dual heterojunctions and each layer is individually connected to a contact pad. The internal photoemission current from both the junctions shows no photocurrent for photons with energy lower than the CBO at the junction. Photons with energy higher than the CBO drive photocurrent which is linearly dependent on two parameters i.e., photon energy/frequency and intensity of light incident as shown in eq.1 and 2. Where the height of CBO determines the threshold frequency for photoexcitation of confined charge carriers above the barrier height and the density of 2-DEG defines the starting point and slope of linearity. Present approach works equivalent to how two equations are used to solve two unknowns. The contribution of light intensity can be reduced by taking difference current which is taken by subtracting the junction currents as given by eq. 3. To reduce the intensity component further to almost zero a suitable multiplication factor (k) is introduced before subtraction as can be seen in eq. 4.

$$I_1 = m_1(\hbar v - \phi_\beta) + n_1 A \quad \ldots \ldots \ldots (1)$$
$$I_2 = m_2(\hbar v - \phi_\beta) + n_2 A \quad \ldots \ldots \ldots (2)$$
$$I_d = (m_1 - m_2)(\hbar v - \phi_\beta) + (n_1 - n_2)A \quad \ldots \ldots \ldots (3)$$
$$I_d = (m_1 - km_2)(\hbar v - \phi_\beta) + (n_1 - kn_2)A \quad \ldots \ldots \ldots (4)$$

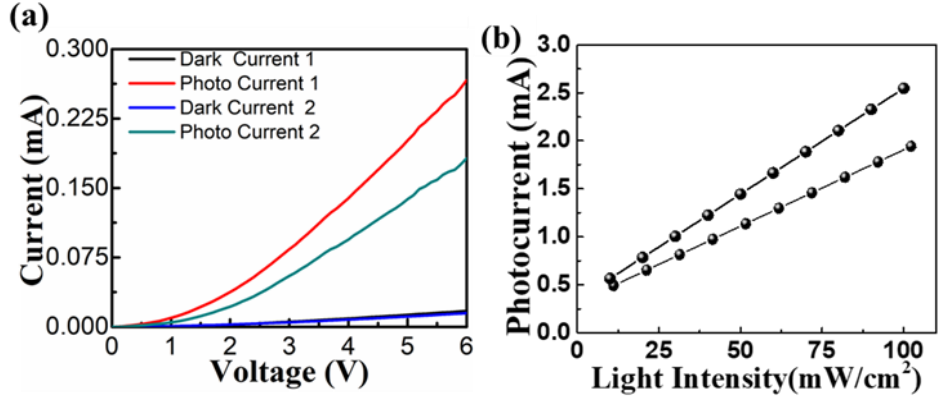

Fig. 3 Opto-electronic response of the device. (a) Dark current and photocurrent of the device for both the junctions with optical excitation from 650 nm wavelength. Dark current and Photocurrent 1 and 2 corresponds to CdZnO-ZnO and ZnO-MgZnO interface respectively (b) Photocurrent of junction 1 and 2 with respect to intensity of the incident light.

Where A is the intensity of light, $\phi_\beta$ is the potential barrier in conduction band at the junction and m and n represents the slopes. The multiplication factor k is judiciously chosen to be $n_1/n_2$, so that the coefficient of intensity (i.e. $n_1 - k*n_2$) becomes zero.

## 3  Device Measurements:

Fig. 2(b) depicts the setup for opto-electronic characterization of the device where the Device under Test (DUT) is placed on top of a vacuum stage and the optical power is irradiated vertically using a continuous wave (CW) laser source. The laser output is connected to a polarization controller via optical fiber through which light is further delivered with the help of optical fiber whose other end is mounted over a 3-axis positioner. The fiber is aligned vertically with respect to the plane of the DUT using a 3-axis positioner as shown in Fig. 2(b) and the whole setup is placed under a microscope stage on an optical table. Light is then illuminated over the DUT and subsequently the optical coupling is optimized by aligning the fiber and the DUT with the aid of micro positioner stage. Opto-electronic response of the device is obtained by providing potential bias to the device through 2-channel Signal Measurement Unit (SMU) while the DUT is illuminated with optical power. Standard BNC cables are connected to the contact pads of the device via electrical probes as shown in the Fig. 3(a) and I-V characteristic of the device is obtained and plotted for with an optical excitation of 650 nm wavelength. Further photocurrents of both the junctions are measured for different light intensities with a fixed voltage bias of 6 V, where the photonic energy is higher than the potential barrier at both the junctions. The photocurrent shows strict linear behavior with respect to the intensity of the light with a slope and responsivity of 1.1 mA/3dB and 25 mA/W respectively, as shown in Fig. 3 (b).

Fig. 4(a) shows the photocurrent response of both the junctions with respect to the energy of incident photon energy ranging from 0.8 to 1.87 eV, where each junction is biased at 6 V. The energy range of the incident photons is well below the corresponding bandgap of any layer, ruling out any contribution of photocurrent due to electron hole pair generation. The photocurrent of both the junctions displays almost linear response for a broad range of frequencies of the incident photons as shown in Fig.4 (a). Junction 1 exhibits higher photocurrent as compared to junction 2 this may be attributed to the fact that it contains higher density of 2-DEG as compared to junction 2 as probability of photoexcitation is directly proportional to the charge density at the interface. The difference between the photocurrent from both the junctions is plotted against photon energy in fig. 4(b) which displays a similar linear dependency over frequency for broad wavelength range of 300 nm.

To take into consideration the effect of noise due to intensity fluctuation from external factors fig. 4(b) is plotted, for a range of light intensity with ±5 dB intensity variations around intensity of 90 mW/cm$^2$. Wavelength sensitivity is calculated by taking the slope from fig. 4(b), where an average change in photocurrent of around 0.96 µA can be seen for corresponding change of 1 nm of wavelength respectively. Thus, wavelength sensitivity of 0.96 µA/nm is reported for a broad wavelength-range of 280 nm. Fig. 4(c) shows difference photocurrent with a multiplication factor of 1.42 for photocurrent in junction 2. According to eq. 4 to remove intensity component from its coefficient (i.e. $n_1 - k*n_2$) should be zero thus multiplication factor (k) is chosen to be $n_1/n_2$, where the individual values of $n_1$ and $n_2$ can be taken from fig. 3(b). To demonstrate intensity noise cancellation in wavelength dependent photocurrent fig. 4(c) is also plotted with error bars. Comparing fig. 4(b) and 4(c) it can be clearly seen that the error margin due to 5 dB fluctuations in intensity is reduced after a multiplication factor of 1.42 is introduced, offering higher resolution. Fig. 4(c) also shows that although, multiplication factor improves the noise performance, it also reduces the wavelength responsivity creating a trade-off. Finally, dynamic response of the proposed device is plotted in fig. 4 (d) where device is excited by a square pulse of laser with a period of 3 µs and 50% duty cycle for 650 and 780 nm with corresponding light intensities of 70 mW/cm$^2$ and 90 mW/cm$^2$ respectively. The current response is plotted in arbitrary units clearly demonstrating wavelength detection capability of the device in parts of visible and IR regime. The speed of the device can be calculated with the help of rise time and fall time of the dynamic response and it comes out to be around 3.6 MHz. The speed as well as responsivity of the device can be enhanced by increasing density of the 2-D confined electron gas. This however will increase the dark current as thermionic emission increases due to thin potential barriers at higher doping, resulting in a tradeoff between speed of the device with dark current. Thus, higher speed would result in larger dark current increasing Noise Equivalent Power (NEP).

# 4 Discussion:

Table 1 provides a comparison between different types of photodetection techniques with potential spectrum detection capabilities. Conventionally used Avalanche Photo Diode (APD) or Photomultiplier Tube (PMT) based detectors exhibits impressive responsivity they however, either hide spectral information of the signal or needs to be couple with bulky frequency scanning components. Limited channels for detection inhibit parallel detection of broadband optical signal along with high spectral resolutions. Another common approach to achieve the same employs dispersion optics which are again either bulky or requires complex designs with extremely low fabrication tolerance. As an alternate semiconductor based absorptive filters have been investigated due to their compactness and on-chip compatibility. The approach utilizes different materials with wavelength selective absorption along with broadband tunable filters. High resolution, broadband spectrum detection with semiconductors can be achieved only by incorporating high quality-factor resonators and multiple layers of different materials respectively. This not only increases the complexity but also leads to excessive production costs. A different approach i.e., superconducting nanowire based photodetection while displaying ultrabroadband functionality, obscure all the spectral information. 2D Materials (2DM) have been recently attracting enormous amount of interest owing to their impressive opto-electronic properties. Riding on the back of ultrahigh responsivity, fast photoresponse and broad detection range, 2DM have become a material of choice for wide range of applications. However, issues such as lack of maturity in large scale growth of high-quality film along with scalable deposition of contacts with highly ohmic character hinders wide scale use of these materials.

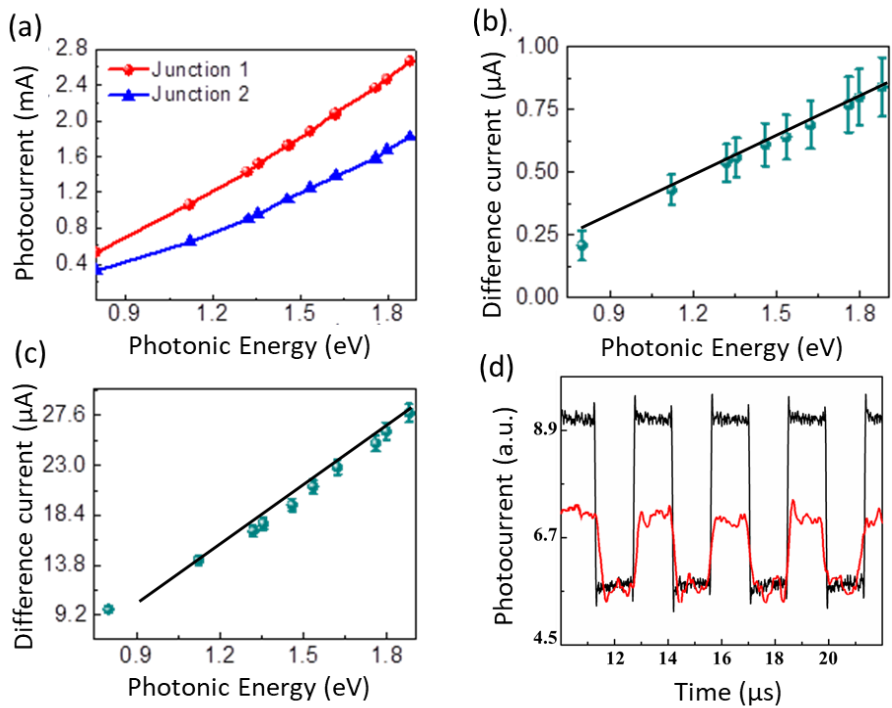

Fig. 4 Frequency response of the device. (a) Junction current for different frequencies with a voltage bias of 6V across the junction. Difference photocurrent response with respect to the photonic energy of the incident light at 6 V bais (b) without multiplication factor. (c) With multiplication factor of 1.42. The error bar represents current fluctuations due to variation in light intensity of ± 5 dB. (d) Dynamic photo response for a laser pulse (650 and 780 nm wavelength) with a frequency of 330 KHz. Trace in red and black denotes dynamic response for 780 nm and 650 nm respectively.

Table 1 Comparison of different wavelength detection techniques [11-15, 38-48]

| Wavelength Detection Techniques | Spectral Bandwidth | Responsivity | Wavelength Resolution | Remarks |
|---|---|---|---|---|
| Avalanche photodiode | ≈ 800 nm | 10 A/W | NA | Excellent responsivity, obscure spectral information |
| Dispersion optics | ≈ 1000 nm | NA | ≈ 1- 2 nm | Either incompatible for on-chip applications or, requires complex fabrication |
| Colloidal Q-Dot | 300 nm | NA | ≈ 1- 2 nm | Incompatible with on-chip photonics |
| Semiconductor nanostructure | 10 nm | NA | 2-3 pm | Excellent wavelength resolution, low fabrication tolerance and narrow bandwidth |
| Superconducting Nanowires | > 1000 nm | NA | Cannot provide spectral information | Ultra-broadband, obscure spectral information |
| 2D Material | ≈ 1000 nm | > 10 A/W | NA | Very high responsivity, growth of large area, high quality, 2D layers is still challenging |
| Present Work | 300 nm | 25 mA/W | Few nanometers | Simple fabrication, broadband low responsivity |

Present work demonstrates, in proof-of-principle, a novel and innovative wavelength detection technique on an on-chip platform with simple fabrication approach. Employing difference in potential barrier and charge carrier concentration in 2-DEG for wavelength specific photocurrent allows bypassing the need for complex geometries and depositing different materials. Different conduction band offset can be used to target different part of the spectrum which can be achieved by changing the MgO and CdO concentration with respect to ZnO. Additionally, presence of a reference junction on the same device provides spatial and temporally coherent photocurrent for performing excellent common mode rejection. It simultaneously makes the device more immune to the noise present in the channel by reducing the intensity dependence of the device. Although the proposed work lacks in speed and responsivity, it is a step forward in the right direction and offers a powerful platform for wavelength detection after further improvements. The major reason for slow speed can be the low mobility of sputtered ZnO films (typically 5-30 cm2/(V.s)) therefore, device would greatly benefit from material engineering for achieving higher mobility. Also thickness of the layers can be reduced for efficient charge transport. Optimized layer thickness along with enhanced mobility will considerably reduce the carrier transit time offering higher speed of the device. Also, surface engineering such as addition of nanorod antennas can reduce reflections and enhance the responsivity of the proposed device. A thorough analysis of the material and design engineering is thus required for further improvement in its responsivity and speed [49, 50].

# 5  Conclusion:

To summarize, wavelength detection capability of the proposed ZnO based dual heterojunction has been experimentally demonstrated for part-visible part-IR region of spectrum over a broad range of frequencies. Internal photoemission via highly dense 2-DEG in ZnO based heterojunction shows linear dependence with respect to frequency and intensity of the incident light. The active region of the device consisting of layers of CdZnO-ZnO-MgZnO, deposited on top of each other incorporates two heterojunctions CdZnO-ZnO and ZnO-MgZnO. Photoemission current via 2-D electron gas formed at both the junctions exhibit linear frequency response. Difference of both the photoemission currents after an appropriate multiplication factor, also displays linear frequency response along with reduced sensitivity to intensity of the light for a broad range of frequencies. This could pave the way for broadband, low noise wavelength detection in optical regime with simple fabrication process. High density of confined electron gas offers enhanced IQE and ensures equal exposure at both the junctions making it ideal platform for on-chip wavelength detection. Wavelength sensitivity of 0.96 µA/nm is reported in the present work. The proposed device can be useful in the fields of optical spectroscopy, sensing, IOT and wavelength demodulation for optical communication.

## Funding:

The authors acknowledge the financial support received from Science and Engineering Research Board (SERB), Government of India with project no. CRG/2020/000144 and from Council of Scientific and Industrial Research (CSIR) with project no. 22(0840)/20/EMR-II. Also, this research work was funded by The Deanship of Scientific Research (DSR) at King Abdulaziz University, Jeddah, Saudi Arabia, under grant no. grant no. (FP-213-42). Therefore, the authors gratefully acknowledge technical and financial support from the Ministry of Education and King Abdulaziz University, Jeddah, Saudi Arabia.

## Acknowledgment:

The authors would like to thank Dr. Suchandan Pal from CSIR-Central Electronics Engineering Research Institute (CSIR-CEERI) Pilani and Hybrid Nanodevice Research Group, IIT Indore for technical support.